\documentclass{kapproc} 

\usepackage{procps} 

\usepackage[dvips]{graphicx}

\upperandlowercase

\kluwerbib 

\begin{document}

\articletitle{Metals in the Neutral Interstellar \\ 
Medium of Starburst Galaxies}
\author{A.~Aloisi$^{1,2}$, T.~M.~Heckman$^{3}$, C.~G.~Hoopes$^{3}$, 
C.~Leitherer$^{1}$, S.~Savaglio$^{3}$, K.~R.~Sembach$^{1}$}
\vspace{0.2cm}
\affil{
$^1$\,Space Telescope Science Institute, 3700 San Martin Drive, 
Baltimore, MD 21218\newline
$^2$\,On assignment from the Space Telescope Division of ESA\newline
$^3$\,Departmnent of Physics and Astronomy, Johns Hopkins University, 
3400 North Charles Street, Baltimore, MD 21218
}

\begin{abstract}
Thanks to their proximity, local starbursts are perfectly suited for 
high-resolution and sensitivity multiwavelength observations aimed to 
test our ideas about star formation, evolution of massive stars, physics 
and chemical evolution of the interstellar medium (ISM). High-resolution 
UV spectroscopy with FUSE and STIS has recently given the possibility to 
characterize in great detail the neutral ISM in local starbursts thanks 
to the presence in this spectral range of many absorption lines from 
ions of the most common heavy elements. Here we present the results for 
two nearby starburst galaxies, I~Zw~18 and NGC~1705, and show how these 
results relate to the star-formation (SF) history and evolutionary state 
of these stellar systems.
\end{abstract}


\section{Introduction}

Star-forming galaxies are an important component of both the low- and 
high-redshift Universe. Star formation is an ongoing phenomenon in 
nearby spiral and irregular galaxies. Starbursts (strong, intense, and 
spatially concentrated episodes of SF) are usually seen in local dwarf 
irregulars and interacting/merging systems. According to the hierarchical
scenario for galaxy formation and evolution (e.g., White \& Rees, 1978), 
merging could have been much more frequent in the past. And indeed in the 
last decade, a population of interacting and star-forming galaxies have 
been discovered at redshifts $z > 1$ in deep fields (e.g., the HDF-N and 
-S), the most famous objects of this type being the so-called Lyman Break 
Galaxies (Steidel et al.~1996).

Thanks to their proximity, local starbursts offer a unique laboratory 
where to perform high resolution and sensitivity multi-wavelength studies 
of all those phenomena related to SF, evolution of massive stars, and 
physics of the ISM (chemical enrichment, gas kinematics, and mixing). 
These detailed studies provide a precious tool with which observations 
of higher redshift star-forming galaxies can be better interpreted (e.g., 
local systems as templates). They also allow us to better understand the 
connection between the low- and high-redshift Universe and the processes 
involved in the origin and evolution of galaxies.

Starbursts are characterized by the presence of a large reservoir of HI 
which is a fundamental component for the onset of the SF process. Recently, 
FUSE and STIS have offered the unique possibility to characterize in great 
detail the neutral ISM in nearby systems by offering access to the plethora 
of absorption lines arising from the most common heavy elements in the UV. 
A multi-component fitting technique is usually applied to absorption lines 
in order to infer the column densities of heavy elements and determine metal 
abundances. The relative abundances of metals which originate in different 
types of stars, are like the fossil record of the past SF history of a 
galaxy, the neutral gas tracing a more ancient past than the ionized gas 
in the H~{\sc ii} regions due to larger timescales of the mixing processes 
that come into play. Different abundance patterns in the neutral ISM are 
thus expected depending on the past SF history of the galaxy.

\section{FUSE Spectra of I~Zw~18}

I~Zw~18 is the star-forming galaxy with the lowest metallicity 
known and has always been regarded as the best candidate for 
a truly ``young'' galaxy in the local universe. It has been 
observed with FUSE for a total of $\sim 90$ (60) ksec in the 
LiF (SiC) channels. The LWRS aperture (30''$\times\,$30'') was 
used to cover the whole body for a resolution of $\sim 35$ km/s 
and a S/N of 7-18 (Aloisi et al.~2003). A multi-component fitting 
technique was applied to infer the column densities of the most 
common heavy elements and determine metal abundances in the 
neutral gas. Our results are reported in Table~1 together with 
the metal abundances in the H\,{\sc ii} regions (Izotov et 
al.~1999; Izotov \& Thuan 1999).

\begin{table}[!ht]
\caption{Interstellar Abundances in I~Zw~18}
\begin{center}
\begin{tabular}{llcccc}
\sphline
Element & Ion & & [X/H]$_{\rm ISM}$ & & [X/H]$_{\rm HII}$\\[3pt] 
\sphline
O  & O~{\sc i}   & & $-2.06 \pm 0.28$ & & $-1.51 \pm 0.04$\\  
Ar & Ar~{\sc i}  & & $-2.27 \pm 0.13$ & & $-1.51 \pm 0.07$\\  
Si & Si~{\sc ii} & & $-2.09 \pm 0.12$ & & $-1.90 \pm 0.33$\\  
N  & N~{\sc i}   & & $-2.88 \pm 0.11$ & & $-2.36 \pm 0.07$\\ 
Fe & Fe~{\sc ii} & & $-1.76 \pm 0.12$ & & $-1.96 \pm 0.09$\\
\sphline
\end{tabular}
\end{center}
\end{table}

It is clear that the neutral gas in I~Zw~18 has already been 
enriched in heavy elements, and is not primordial in nature. 
The $\alpha$ elements are several times lower in the H\,{\sc i} 
than in the H\,{\sc ii} gas, while Fe is the same. The Fe behavior 
suggests that some old SF is required for the metal enrichment of 
the H\,{\sc i} (Fe is mostly produced by SNe\,Ia on time 
scales $> 1$ Gyr). The relative metal content in $\alpha$ elements
(produced by SNe\,II on time scales $< 50$ Myr) and N (released 
on timescales $> 300$ Myr) between neutral and ionized gas suggests 
that the H\,{\sc ii} regions have been additionally enriched by 
more recent SF.

\section{FUSE and STIS Spectra of NGC~1705}

There are some caveats to address in order to be sure that what 
we really measure is the ISM metal abundance. First of all, there
is saturation, especially for O\,{\sc i}. Hidden saturation of 
unresolved multiple components can also bring to erroneous estimates 
of the column density. Ionization constitutes another type of 
uncertainty. Abundances are derived by assuming that the primary 
ionization state in the neutral gas is representative of the total 
amount of a certain element and this is probably the case in ISM 
studies. However, we can have contamination by ionized gas laying 
along the line of sight. Furthermore, Ar\,{\sc i} and N\,{\sc i} 
should be well coupled with H\,{\sc i} and O\,{\sc i} due to similar 
ionization potentials. However, they could be found in higher 
percentage in their ionized state due to larger cross-sections for 
photoionization. Finally, depletion is important. Some elements 
could be more easily locked into dust grains than others (e.g., 
Fe compared to O), thus altering the relative abundances produced 
by a certain SF history.\looseness=-2

A wonderful dataset where to address all these issues 
is represented by the FUSE and STIS Echelle 
spectra (900-3100 \AA) of NGC~1705, one of the brightest dwarf 
starburst galaxies in the nearby Universe. The FUSE data were 
taken by centering the SSC in the LWRS aperture for a total of 
$\sim 21$ ksec, a resolution of $\sim 30$ km/s and a S/N of 
10-16 (Heckman et al.~2001). The STIS 
echelle data were taken with the 0.2''$\times\,$0.2'' aperture 
centered on the SSC for a total of 10 HST orbits, a resolution of 
$\sim 15$ km/s, and a S/N of 10-20 
(V\'azquez et al.~2004). We measured the column density of 
many ions in the FUSE and STIS spectra of NGC~1705 with the 
line-profile fitting and inferred the ISM metal abundances. We 
found consistency in the measurements performed on those ions 
detected in both spectra. Thus, the FUV light is 
dominated by the SSC. The low-ionization absorption lines have a 
mean radial velocity of about 590 km/s. However, two components 
were detected for selected ions in the higher-resolution STIS 
data. One component is at the same radial velocity of the stars 
in the SSC ($v = 618$ km/s) as inferred by the stellar C\,{\sc iii} 
line at 1175 \AA, and the second at the velocity of the warm photoionized 
gas ($v = 580$ km/s) as detected in absorption through C\,{\sc iii} 
or N\,{\sc ii} and confirmed by nebular emission lines in the optical. 
The use of the abundances from the total (neutral $+$ ionized) absorbing 
gas would thus be misleading for the derivation of the metal content in 
the neutral ISM of NGC~1705. In Table~2 we report the abundances as 
inferred by both the total column densities of the ions (column 3) and 
the column densities of the absorbing component at rest with the stars 
in the SSC (column 4). 
The latter have to be compared with the H\,{\sc ii} region abundances
(Lee \& Skillman 2004).\looseness=-2

\begin{table}[!ht]
\caption{Interstellar Abundances in NGC~1705}
\begin{center}
\begin{tabular}{llcccccc}
\sphline
Element & Ion & & [X/H]$_{\rm ISM,total}$ & & [X/H]$_{\rm ISM,neutral}$ & 
& [X/H]$_{\rm HII}$\\[3pt] 
\sphline
O  & O~{\sc i}   & & $-1.19 \pm 0.01$ & &        ...       & & $-0.48 \pm 0.05$\\  
Ar & Ar~{\sc i}  & & $-1.11 \pm 0.04$ & &        ...       & & $-0.61 \pm 0.10$\\  
Si & Si~{\sc ii} & & $-0.90 \pm 0.01$ & &        ...       & &        ...      \\  
Mg & Mg~{\sc ii} & & $-1.41 \pm 0.12$ & &        ...       & &        ...      \\ 
Al & Al~{\sc ii} & & $-1.14 \pm 0.04$ & & $-1.36 \pm 0.05$ & &        ...      \\ 
N  & N~{\sc i}   & & $-1.79 \pm 0.03$ & & $-2.29 \pm 0.06$ & & $-1.51 \pm 0.08$\\
Fe & Fe~{\sc ii} & & $-0.86 \pm 0.03$ & & $-1.29 \pm 0.03$ & &        ...      \\
\sphline
\end{tabular}
\end{center}
\end{table}

\section{Conclusions}

The offset in metal content between neutral ISM and H\,{\sc ii} regions 
in local starburst galaxies is probably one of the unexpected great results 
of the last years. However, this area of research is still pretty new 
and many unknowns and uncertainties still affect the interpretation of the 
data. It is thus premature to draw conclusions and more targets with a data 
quality similar to that of NGC~1705, still need to be investigated before 
having all the pieces of this intriguing puzzle put together.

\begin{chapthebibliography}{1}

Aloisi, A., Savaglio, S., Heckman, T.~M., Hoopes, C.~G., 
Leitherer, C., \& Sembach, K.~R.~2003, ApJ, 595, 760

Heckman T.~M., Sembach, K.~R., Meurer, G.~R., Strickland, 
D.~K., Martin, C.~L., Calzetti, D., \& Leitherer, C.~2001,
ApJ, 554, 1021

Izotov, Y.~I., Chaffee, F.~H., Foltz, C.~B., Green, R.~F., 
Guseva, N.~G., \& Thuan, T.~X.~1999, ApJ, 527, 757

Izotov, Y.~I., \& Thuan, T.~X.~1999, ApJ, 511, 639

Lee, H., \& Skillman, E.~D.~2004, ApJ, in press, astro-ph/0406571

Steidel, C.~C., Giavalisco, M., Pettini, M., Dickinson, M.,
Adelberger, K.~L.~1996, ApJ, 462, L17

V\'azquez, G.~A., Leitherer, C., Heckman, T.~M., Lennon, 
D.~J., de Mello, D.~F., Meurer, G.~R., \& Martin, C.~L.~2004,
ApJ, 600, 162

White, S.~D.~M., \& Rees, M.~J.~1978, MNRAS, 183, 341

\end{chapthebibliography}

\end{document}